# Characterizing Aqueous Foams by In-situ Viscosity Measurement in a Foam Column

Wei Yu[1,*], Jack Hau Yung Lo[1,*], Mazen Yousef Kanj[1]

[1]Center for Integrative Petroleum Research (CIPR), College of Petroleum Engineering & Geosciences, King Fahd University of Petroleum & Minerals, Dhahran 31261, Saudi Arabia

[*]Corresponding authors: wei.yu@kfupm.edu.sa; hau.yung@kfupm.edu.sa

**Abstract**

Foam characterization is essential in many applications of foams, such as cleaning, food processing, cosmetics, and oil production, due to these applications' diversified requirements. The standard characterization method, the foam column test, cannot provide sufficient information for in-depth studies. Hence, there have been many studies that incorporated different characterization methods into the standard test. It should be enlightening and feasible to measure the foam viscosity, which is both of practical and fundamental interest, during the foam column test, but it has never been done before. Here, we demonstrate a method to characterize aqueous foams and their aging behaviors with simultaneous measurement of foam viscosity and foam height. Using a vibration viscometer, we integrate foam column experiments with in-situ foam viscosity measurements. We studied the correlation among the foam structure, foam height, and foam viscosity during the foam decay process. We found a drastic decrease in foam viscosity in the early foam decay while the foam height remained unchanged, which is explained by coarsening. This method is much more sensitive and time-efficient than conventional foam-height-based methods by comparing the half-life. This method successfully characterizes the stability of foams made of various combinations of surfactants and gases.

**Key words:** foam characterization; foam againg; foam stability; viscosity measurement; foam column; foam height; coarsening; $CO_2$ foam

**Introduction**

Aqueous foams are concentrated dispersions of gas bubbles in aqueous solutions. Foams are ubiquitous in everyday life and have been of extensive interest in many industrial applications, such as cleaning, the food industry, cosmetics, and oil production [1,2]. Foams can be conveniently created by dispersing gases into surfactant solutions, but they are thermodynamically unstable due to high surface free energy [3]. Foam lifetime can vary in a broad spectrum from several seconds (e.g., beer foam) to minutes (e.g., soap foam) or even months (e.g., nanoparticle-stabilized foam) due to various components and structures [1,4]. Characterizing foam properties has always been important due to the versatile requirement in different applications.

Three mechanisms are responsible for foam decay and its microscopic structural change [2,3]. The first mechanism, drainage, is the liquid flow between bubbles driven by gravity and capillarity [5]. The second mechanism, coarsening, is due to the gas diffusion between bubbles driven by curvature differences, resulting in the increase of average bubble size [6]. The third one, coalescence, is due to the rupture of liquid films [7]. However, collecting all information about the microscopic structural change and then deducing the macroscopic effect by theoretical models is challenging. In applications, it is more common to characterize the foam properties by macroscopic properties of the foam, such as the well-known foam column test (e.g., ASTM D1173 and ISO 696).

In a typical foam column experiment, foam is generated in a glass column by introducing a gas into a surfactant solution using methods like mechanical stirring, pouring (Ross-Miles method [8]), shaking (Bartsch method [9]), or bubbling [10,11]. The maximum foam height is used as a measure of foamability. As foam decays, the foam volume decreases. Therefore, the rate of fall of foam height is used to characterize foam stability. To exploit more information, different variations of foam column experiments have been proposed, including using a high-resolution camera to measure the microscopic structure of foam, using optical or electrical detectors to measure for example the liquid distribution and liquid fraction, etc. [10,12,13].

To the best of our knowledge, an in-situ measurement of viscosity in a foam column has not been reported before[14]. Viscosity is one of the most intriguing properties of foams. Foams exhibt much higher viscosity than either the gas or the liquid phase from which it is comprised. This feature is useful in many applications, such as increasing the displacement efficiency in oil recovery [15,16] and texturizing food and cosmetics [1]. Foams are viscoelastic and foam viscosity is equal to the ratio of the loss modulus to the oscillation frequency in rheology. Foam viscosity is determined by the interfacial properties of foam components and the microscopic structure of foams [17–19]. As the foam ages or decays, its microscopic structure and hence foam viscosity changes over time. Cohen-Addad et al. studied the influence of the coarsening on the complex shear modulus of aqueous foam and showed a decrease of loss modulus as the foam coarsens[20]. Soller and Koehler studied the influence of drainage by performing rheological measurements of aqueous foam, skillfully controlling the liquid content and bubble size by liquid perfusion and continuous bubble generation [18].

Therefore, the significance of the in-situ measurement of viscosity is twofold. First, the foam viscosity directly impacts the foam's performance in various applications. Second, the foam viscosity is correlatable to its time-dependent decay behavior. Previous measurements of foam viscosity are mainly carried out using rheometers that are incompatible with the foam column experiments [21,22]. In rheometers, the sample liquid (foam) is confined in a thin gap, which is ~1 mm thick typically (see SI). The thin gap of the rheometer turns the foam into a quasi-2D system and thus affects the foam decay process. Another less critical issue is that the sample holders of rheometers are usually non-transparent (except for tailor-made setups), preventing the direct observation of foam that is required for the foam column test.

This paper describes a novel method for characterizing the decay of aqueous foam by the simultaneous measurement of foam viscosity and foam height. We integrate foam column experiments with in-situ foam viscosity measurements using a vibration viscometer. Without disturbing the normal structure and decay of the foam column, the proposed method allows for a time-efficient and sensitive measurement of foam viscosity. The viscosity measurements reveal an early-stage decay of foam dominated by coarsening, which is not detectable from typical foam height measurements. We utilize this method to analyze foams made of various combinations of surfactant solutions and gases.

**Methods**

**Materials**

Three typical commercial surfactants were used as the foaming agents for demonstration: sodium dodecylsulfate (SDS), 3-[Dimethyl(tetradecyl)ammonio]-1-propanesulfonate (SB3-14), and Triton X-100, which are anionic, zwitterionic, and nonionic, respectively. All surfactants were purchased from Sigma-Aldrich. The surfactant solutions were prepared using deionized (DI) water with a concentration higher than their respective critical micelle concentration (CMC). At concentrations higher than CMC, it is known that the coalescence between bubbles rarely occurs [5,23,24]. High-purity $N_2$ (99.9%) and $CO_2$ (99.9%) from Saudi Industrial Gas Company Ltd. (Dammam, Saudi Arabia) were used as the gas phases. $CO_2$ was mixed with $N_2$ to tune the gas solubility in the liquid phase.

**Apparatus**

The schematic of the apparatus for in-situ foam viscosity measurement in a foam column is shown in **Fig. 1**. Foam is generated by flowing gases through surfactant solutions from a sintered glass frit at the bottom of a glass column. The glass frit has a pore size of 10-15 μm, generating bubbles with an initial size of around 200 μm under our testing condition. The glass column has a volume of 150 mL, an inner diameter of 70 mm, and a height of 60 mm. The container size is large enough to avoid affecting the vibration viscosity, as the penetration depth of the oscillation (i.e., decay length of the amplitude) is calculated to be less than 8mm. Gas is injected into the surfactant solution at a constant flow rate regulated by a mass flow controller (GE50A, MKS Instruments). Between the frit and mass flow controller, there is a valve and a gas mixer. The valve is used to prevent the backflow of surfactant solution. The mixer with a volume of 600 mL is used to stabilize the gas flow rate and to mix the $N_2/CO_2$ stream better. A vibrational viscometer (SV-10, A&D Company) controlled by a computer is used to measure the viscosity of the decaying foam. The oscillator frequency is 30 Hz, and the vibration amplitude is 400 μm. The gold sensor plate has a diameter of 13 mm with a surface area of 133 $mm^2$ and a thickness of 0.5 mm. The temperature of the foam is measured simultaneously with foam viscosity. All experiments were performed at room temperature (22 ± 1 °C) and ambient pressure. A high-resolution camera (MG-507C, Allied Vision) was used to capture the time-lapse images of the foam. We also cross-checked the validity of the vibrational viscometer with a rheometer (MCR 702, Anton Paar).

**Procedure of measurement**

Before each test, the glass column and the glass frit were rinsed with isopropanol and water each and then dried in the oven at 80 °C for 2 hours. The viscometer's vibrating plates (sensors) were pre-hung inside the column at a specific height aligned with the column's geometric center. The front and side view of the set-up are shown in **Fig. 1a**. For foam generation, we first removed the residual air by injecting the gas through the porous glass frit at a rate of 100 mL/min for 3 min. Next, we gently poured 30 mL of surfactant solution into the glass column and let the gas flow through the porous frit into the solution for 90 s at a rate of 80 mL/min. The resultant monodisperse bubbles accumulated in the glass column, forming the final foams that fully covered the viscometer's sensor plates. Lastly, we quickly stopped the gas injection by closing the valve and started the measurements. We set time $t = 0$ as the start of measurement. A photo of a typical measurement at the time $t = 0$ is shown in **Fig. 1b**.

**Working principle of a vibrational viscometer**

The working principle of a vibrational viscometer is based on the vibration of thin plates that are immersed in the testing medium[25], as shown in **Fig. 1a**. The plates vibrate sinusoidally at a fixed frequency and amplitude. The oscillation direction is parallel to the plates' surfaces, as indicated by the arrows in **Fig. 1a**. The vibrating plates produce a shear wave. One can show that the corresponding viscous frictional force is $f \sim \omega^{3/2} A\sqrt{\rho \mu}$, where $\omega$ is the angular oscillation frequency, $A$ is the oscillation amplitude, $A$ is the density of the medium, and $\mu$ is the viscosity (**see SI for the derivation**). Both the frequency $\omega$ and amplitude $A$ are constant. The density $\rho$ is measured by a weighing balance. Therefore, the viscosity can be deduced by measuring the dissipation or the impedance from the electronic system. Compared to other types of viscometers, the vibrational viscometer provides for continuous and non-destructive measurement of the essential properties of complex fluids, such as the coagulation of egg albumen and the cloud point of surfactants[26,27]. Based on the diameter of the sensor plates (13 mm) and bubbles (200-600 μm), we estimate that 1800-17000 bubbles are in contact with the vibrating plates at any moment. Thus, the sample size is sufficiently large. We also assume a no-slip boundary condition at the sensor plates.

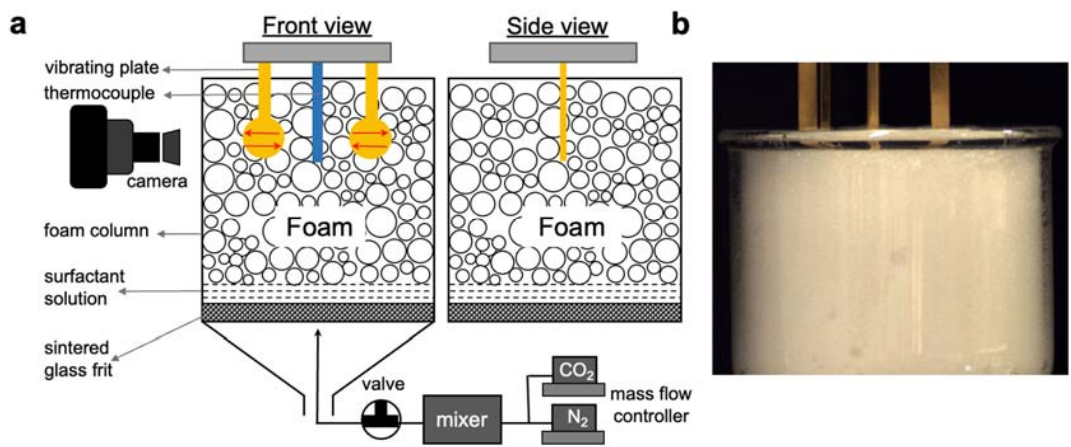

**Figure 1.** Experimental setup. (a) Schematic of the apparatus. (b) Photo of the set-up and foam.

**Results and Discussion**

**Method verification**

To verify the reliability of our measurements using the vibrational viscometer, the viscosity of 10 mM SDS/air foam was measured by both the vibrational viscometer and a rheometer. In our set-up, the angular oscillation frequency is 188 s$^{-1}$, and the shear strain is ~9% (see **SI**). The viscosity of the foam is independent of the strain and increases linearly with the frequency, as shown by the strain sweep and frequency sweep measurements of the rheometer (**Fig. S2**). The results of the two methods show good agreement with each other (see **SI**).

We also verify that our experiments are highly reproducible by repeating an experiment 5 times, showing excellent reproducibility with an error < 1.5%, as shown in **Fig. S3** in the supplementary. We conducted the intermittent on-off test, as shown in **Fig. 2**, to demonstrate that the vibration of the probes does not disturb the foam decay. The test consists of two trials. In the first trial, the viscometer is continuously turned on as normal. In the second trial, the viscometer, so as the vibration, was switched on and off intermittently. The measured viscosity of both trials is plotted in **Fig. 2**. The two sets of data are highly aligned, suggesting that the vibration of sensor plates does not affect the foam.

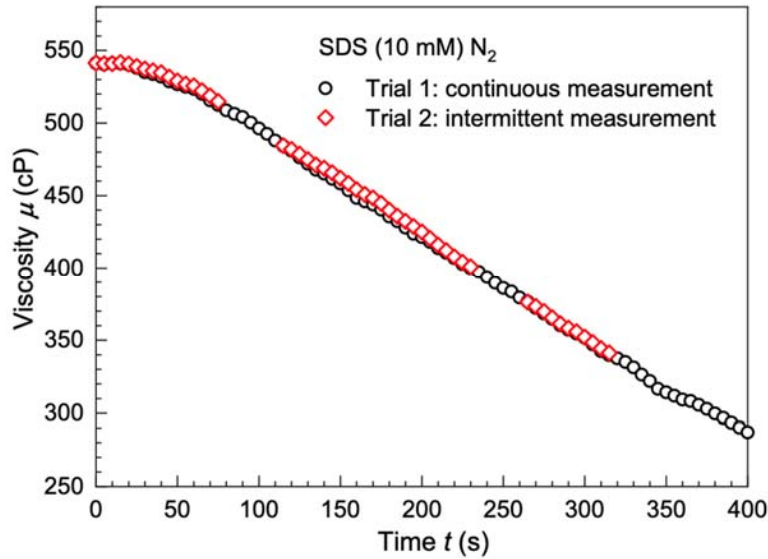

**Figure 2.** Intermittent measurement of foam viscosity using a vibrational viscometer in a foam column. Foam viscosity was measured continuously in trial 1. Foam viscosity was measured intermittently by switching on and off the viscometer.

**Typical measurement results and analysis**

The results of an experiment measuring viscosity and foam height simultaneously over time are presented in **Fig. 3a**. The foam is made of 10 mM SDS aqueous solution and pure $N_2$ gas. The foam viscosity was measured in-situ in the foam column by the vibrational viscometer. During stage I (time $t < 600$ s), the foam viscosity (black circles) decreases significantly, decreasing by 60%, but the foam height remains almost constant ($\Delta h \sim 0$, red circles). This demonstrates that the viscosity measurement reflects changes in foam structure that the typical foam height measurement could not, and it provides a more sensitive detection of foam decay.

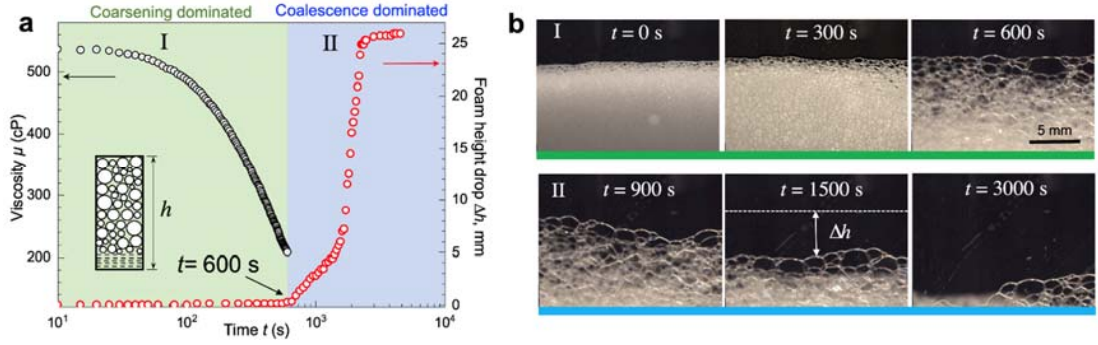

**Figure 3**. **Measurement results.** (a) Viscosity $\mu$ (black) and the decrease in foam height $\Delta h$ (red), defined as $\Delta h(t)=h(0)-h(t)$, versus time $t$. During stage I ($t<600$ s), the foam viscosity decreases significantly, decreasing by 60%, but the foam height remains almost constant. Coarsening dominates the foam decay. During stage II ($t > 600$ s), the foam height decreases significantly due to coalescence. (b) Photos of the foam showing that during stage I, average bubble size increases significantly without a noticeable change in the foam height. During stage II, the foam height decreases significantly. The foam is made of SDS (10 mM) and $N_2$ gas. The scale bar is 5 mm.

The rapid decrease in viscosity during stage I is mainly due to coarsening – By looking at the photos of the foam (**Fig. 3b**), we find that the bubble size increases gradually with time due to gas diffusion. As the average bubble diameter $d$ increases, it is known that the viscosity of foam $\mu$ will decrease [19,20,28]. Two microscopic images of foams at $t=10$ and $t= 600$ s are shown in **Fig. 4a & b**. From the microscopic images, we measured the average bubble diameter versus time as shown in the inset of Fig.4c. The sample size for calculating the average diameter is between 60-80 bubbles. By combining the data of viscosity and average diameter over time, a plot of the foam viscosity $\mu$ versus the average bubble diameter $d$ is shown in **Fig. 4c** in log-log scale, and the best fit is $\mu \propto 1/d^{0.94}$. This inverse relationship agrees with the model proposed by Buzza et al. that $\mu \propto \mu_d/d$, where $\mu_d$ is the surface dilational viscosity [19]. The foams are assumed to be incompressible because typically their bulk modulus is several orders of magnitude larger than their shear modulus[29].

We can exclude the possibility that a decrease in viscosity during stage I is caused by coalescence. Previous experimental studies have shown that coalescence rarely occurs under similar experimental conditions [5,23,24]. Evidently, we also found no sign of coalescence in stage 1 from our recorded movie of the bubble over time.

We also find that the contribution of drainage to the change in viscosity is insignificant in our case. The liquid fraction of the foam at $t= 0$ s and $t= 600$ s was measured to be 0.6% and 0.4%, respectively, by the weighing method. In general, the change in liquid fraction is primarily contributed by the drainage in Plateau borders, with the contribution of film drainage being secondary. Soller et al. have proposed a model of first-order approximation to describe the film thickness as a function of liquid fraction[17,18]. In this model, $h = h_0(1 + h'\epsilon)$, where $h'$ is a fitting parameter, known as the film-swelling parameter, $\epsilon$ is the liquid fraction, and $h_0$ is the equilibrium thickness. Given that the change in the liquid fraction is $\Delta\epsilon$=0.002 in our case and the best-fit value of $h'$ is ~30 for SDS foam[17,18], we estimate that $h'\Delta\epsilon$~0.05. Therefore, the change in film thickness due to drainage is minimal (<5%), and so is its contribution to the change in viscosity.

During stage II (time $t > 600$ s), the foam height starts to decrease significantly due to the foam film rupture in the upper layers (bubble burst). The collapse of foam causes a decrease in the foam volume and thus the foam height. We did not show the viscosity in stage II because the sensors were no longer fully immersed in the foam, which led to inaccurate results. We plan to address this issue and improve our setup in the near future.

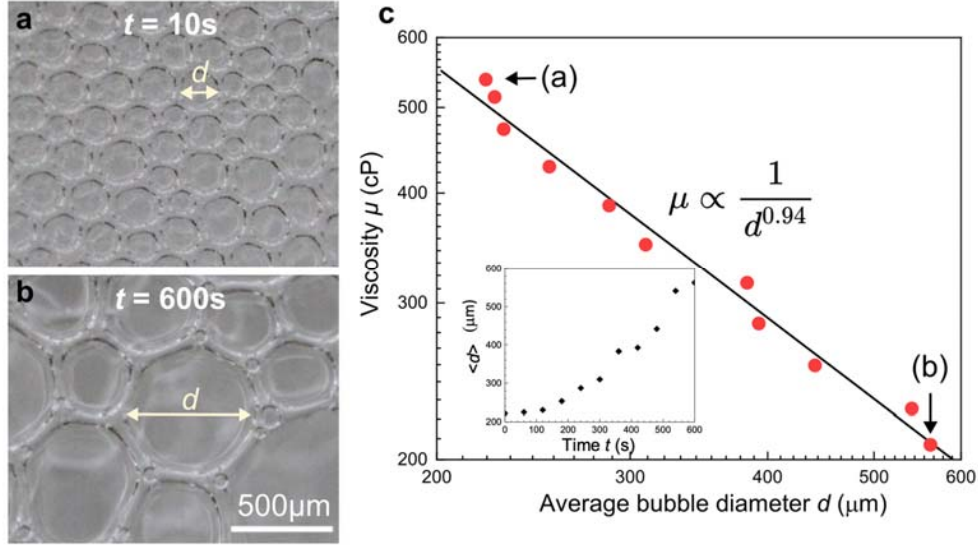

**Figure 4**. **Relation between viscosity and bubble size.** (a, b) Microscopic images of foam at time $t =$ 0 and 600 s. We measured the diameters of 60 to 80 bubbles, $d_i$, and calculated the average bubble diameter $d$ at different times. (c) The plot of foam viscosity $\mu$ as a function of average bubble diameter $d$ in log-log scale. The best fit gives $\mu \propto 1/d^{0.94}$. The foam is made of SDS (10 mM) and $N_2$ gas.

**Results of various foams**

Next, we adopted this method to measure the viscosity of foams made of different surfactant solutions, including SB3-14 (50 mM), SDS (70 mM), and Triton X-100 (80 mM), which are zwitterionic, anionic, and nonionic respectively and $N_2$ gas. As shown in **Fig. 5a**, the viscosity of all three foams decreases with time, as expected. However, they show different initial values and rates of reduction. To better understand the rates of viscosity reduction response, we also normalize the data by the initial viscosity at $t = 0$, as shown in **Fig. 5b**.

Firstly, all three foams have the same initial foam height under the same foaming procedure (see section 2.3). That means all the foaming agents have the same foamability but different initial foam viscosity (**Fig. 5a**): SDS foam has the highest initial viscosity, followed by SB3-14 and Triton X-100 (602 vs. 564 vs. 474 cP). This confirms that foam viscosity and foam height (volume fraction of bubbles) are different intrinsic properties, as expected. Moreover, the fact that all three foams have the same initial height but different viscosity implies that their initial foam structures are different. Therefore, measuring foam viscosity can provide more information about the initial foam structures beyond foam height.

Secondly, from the decreasing rates of viscosity (**Fig. 5b**, lines), the results show that the SB3-14 foam is more stable than either SDS or Triton X-100 foam. This agrees with the foam height responses as well as the literature [30,31]. In the foam column test, it is common to compare the foam stability quantitatively by the half-life. We list the half-life of the three foams obtained by the viscosity and height measurements in **Table 1**. The half-life in viscosity measurements ranges from 6 to 10 minutes, while the half-life in the height measurements ranges from 26 to 152 minutes, showing that we can deduce the stability much faster (up to 15×) by using the new method. Alternatively, instead of the half-life, if we compare the change after 1 min (as per the ASTM D1173), the advantage of the viscosity method is even more obvious.

Finally, we used the viscosity-based method to characterize foams made of various gas compositions. We prepared SDS foams with mixtures of $N_2$ and $CO_2$ gas. The concentration of $CO_2$ varies from 0 to 20%. Our measurements show that the foam viscosity decreases more rapidly as the $CO_2$ concentration increases, as shown in **Fig. 5c**. This is because the water solubility of $CO_2$ is ~100 times higher than $N_2$. As such, an increase in $CO_2$ concentrations will increase the speed of diffusion and coarsening [32]. Our previous analysis has already shown that the decrease in viscosity is mainly due to coarsening (**Fig. 4**). Therefore, the results agree with our expectations that, as more $CO_2$ is mixed with $N_2$, the foam becomes less stable.

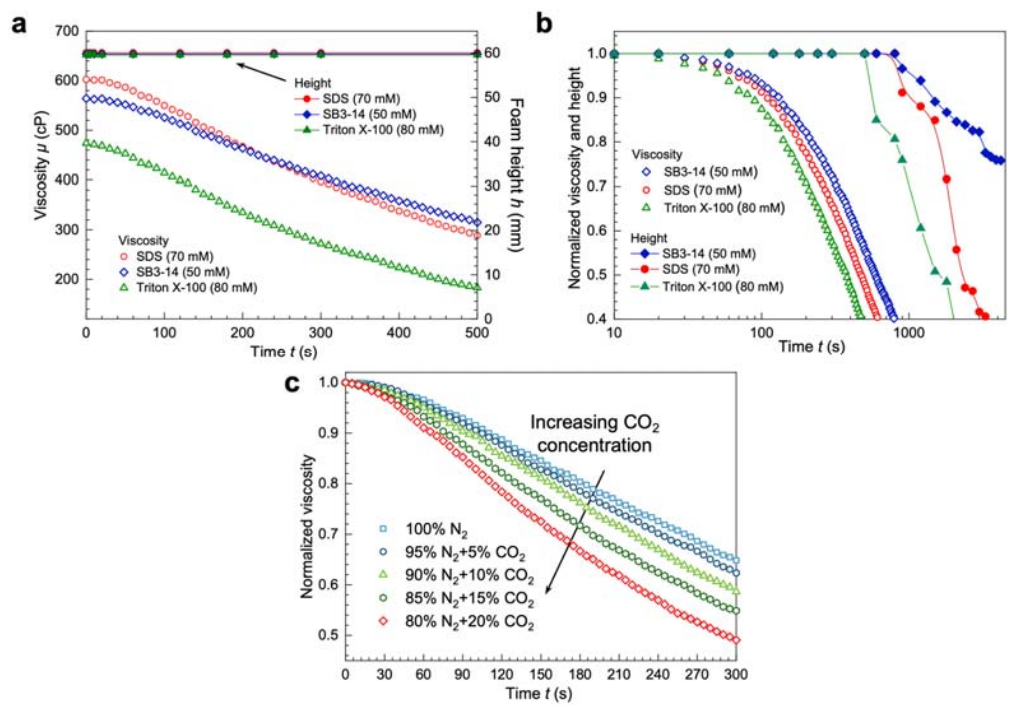

**Figure. 5. Comparing foams made of different combinations of surfactants and gases**. (a) Viscosity and foam height versus time for three different types of foam made of $N_2$ and one of three surfactants SB3-14 (50 mM), SDS (70 mM), or Triton X-100 (80 mM), respectively. (b) Normalized foam viscosity and foam height versus time. (c) Normalized viscosity over time for foam made of SDS (10mM) and increasing ratio of $CO_2$ in $N_2/CO_2$ mixture gas.

**Table 1.** Half-life of SB3-14, SDS, and Triton X-100 foams measured by foam viscosity and foam height.

| Types of foam | Half-life (minutes) | |
| --- | --- | --- |
| | By foam viscosity | By foam height |
| SB3-14 | 10 | 152 |
| SDS | 8 | 38 |
| Triton X-100 | 6 | 26 |

**Conclusions**

We developed a novel method to characterize aqueous foams with simultaneous measurement of foam viscosity and foam height. Using a vibration viscometer, we integrated foam column tests with in-situ foam viscosity measurements. The validity of the vibration viscometer was cross-checked by a rheometer. We also proved that the vibration itself does not disturb the foam structure. There were many previous studies to incorporate different characterization methods into the widely-accepted foam column test. Nevertheless, to the best of our knowledge, this study is the first attempt to incorporate viscosity measurement, which is both of practical and fundamental interest.

We revealed a drastic decrease in foam viscosity in the early stage of foam decay while the foam height remained unchanged. In other words, the viscosity measurement reflects changes in foam structure that the typical foam height measurement could not. We also showed that the foam decay process consists of two stages that are separately dominated by coarsening and coalescence. Coarsening caused a drastic decrease in foam viscosity, and it agrees with the previous model that the viscosity is inversely proportional to the bubble size and proportional to the surface dilational viscosity. In previous studies of foam viscosity during aging, the foam is typically confined within a thin gap that limits the number of layers of bubbles. The rearrangement and aging of foams in such confined spaces are likely different from that in the foam column due to finite-size effects. We also characterized foams made of various combinations of surfactants and gases. The proposed method has demonstrated excellent capabilities in analyzing foam stability compared to the conventional foam-height-based method. The half-life analysis of foam viscosity allowed deducing the foam stability nearly 15 times faster than the conventional counterpart (Table 1). While both foam viscosity and foam height serve to characterize foam stability, they represent distinct aging mechanisms; thus, measuring both parameters provides a more comprehensive foam assessment.

Foam viscosity measurement is important to a variety of foam applications, where foam viscosity is commonly used as a measure of foam strength, e.g., in enhanced oil recovery and mineral flotation. Our method can be applied to characterize more complex foam systems, such as foams stabilized by polymers, emulsions, nanoparticles[33,34], and capillary foams[35]. Further, the characterization of foams made of different gases has attracted increasing attention due to the efficacy of foams in underground gas storage, e.g., $CO_2$, $CH_4$, and $H_2$. This study shed light on understanding the stabilization mechanism and the practical performance of foams.

## Supporting Information

Additional details to the working principles of the experimental system

## Acknowledgements

We thank Drs. Zhengwei Pan and Yafei Chen for their help in experiments and fruitful discussions, and Mr. Jafar Sadeq Al Hamad for his assistance in using the rheometer. Jack Lo acknowledges the financial support from the Agreement TK1914385 between KFUPM, DTVC, and CUHK (project ID 7010538) and RMGS 7010538 (project ID 8601417).

## Author Contributions

**Wei Yu** conceptualized the idea, conducted the experiments, and wrote the manuscript. **Jack Lo** conceptualized the idea, supported the experimental study, and wrote the manuscript. **Mazen Kanj** recommended experiments and revised the manuscript.

## Conflict of interests

The authors declare no competing financial interest.

## References

(1) Hill, C.; Eastoe, J. Foams: From Nature to Industry. *Adv Colloid Interface Sci* **2017**, *247*, 496–513. https://doi.org/10.1016/J.CIS.2017.05.013.

(2) Stevenson, P. *Foam Engineering: Fundamentals and Applications*; Paul Stevenson, Ed.; A John Wiley & Sons, Ltd., Publication, 2012.

(3) Cohen-Addad, S.; Höhler, R.; Pitois, O. Flow in Foams and Flowing Foams. *Annu Rev Fluid Mech* **2013**, *45*, 241–267. https://doi.org/https://doi.org/10.1146/annurev-fluid-011212-140634.

(4) Binks, B. P.; Murakami, R. Phase Inversion of Particle-Stabilized Materials from Foams to Dry Water. *Nat Mater* **2006**, *5* (11), 865–869. https://doi.org/10.1038/nmat1757.

(5) Koehler, S. A.; Hilgenfeldt, S.; Weeks, E. R.; Stone, H. A. Drainage of Single Plateau Borders: Direct Observation of Rigid and Mobile Interfaces. *Phys Rev E* **2002**, *66* (4), 4. https://doi.org/10.1103/PhysRevE.66.040601.

(6) Taccoen, N.; Dollet, B.; Baroud, C. N. Order to Disorder Transition in a Coarsening Two-Dimensional Foam. *Phys Rev Lett* **2019**, *123* (23). https://doi.org/10.1103/PhysRevLett.123.238006.

(7) Forel, E.; Dollet, B.; Langevin, D.; Rio, E. Coalescence in Two-Dimensional Foams: A Purely Statistical Process Dependent on Film Area. *Phys Rev Lett* **2019**, *122* (8). https://doi.org/10.1103/PhysRevLett.122.088002.

(8) Rosen, M. J.; Solash, J. Factors Affecting Initial Foam Height in the Ross-Miles Foam Test. *J Am Oil Chem Soc* **1969**, *46* (8), 399–402. https://doi.org/10.1007/BF02545623.

(9) Petkova, B.; Tcholakova, S.; Denkov, N. Foamability of Surfactant Solutions: Interplay between Adsorption and Hydrodynamic Conditions. *Colloids Surf A Physicochem Eng Asp* **2021**, *626*, 127009. https://doi.org/https://doi.org/10.1016/j.colsurfa.2021.127009.

(10) Lunkenheimer, K.; Malysa, K.; Winsel, K.; Geggel, K.; Siegel, S. Novel Method and Parameters for Testing and Characterization of Foam Stability. *Langmuir* **2010**, *26* (6), 3883–3888. https://doi.org/https://doi.org/10.1021/la9035002.

(11) William G. Loisel. Device for the Characterization of the Foaming Properties of a Product Which Is at Least Partially Soluable. US5465610A, 1995. https://patents.google.com/patent/US5465610A/en.

(12) Wang, Q.; Wang, D.; Shen, Y.; Wang, H.; Xu, C. Influence of Polymers on Dust-Related Foam Properties of Sodium Dodecyl Benzene Sulfonate with Foamscan. *J Dispers Sci Technol* **2017**, *38* (12), 1726–1731. https://doi.org/10.1080/01932691.2016.1278550.

(13) Monnereau, C.; Vignes-Adler, M. Optical Tomography of Real Three-Dimensional Foams. *J Colloid Interface Sci* **1998**, *202* (1), 45–53. https://doi.org/10.1006/JCIS.1998.5437.

(14) Wei Yu; Jack H.Y. Lo; Zhengwei Pan; Mazen Yousef Kanj. Method for Characterizing and Evaluating Foamability and Foam Stability Using a Vibrational Viscometer. Application Serial No: 63/399,833, August 22, 2022.

(15) Yu, W.; Kanj, M. Y. Review of Foam Stability in Porous Media: The Effect of Coarsening. *J Pet Sci Eng* **2021**, 109698. https://doi.org/https://doi.org/10.1016/j.petrol.2021.109698.

(16) Omirbekov, S.; Davarzani, H.; Colombano, S.; Ahmadi-Senichault, A. Experimental and Numerical Upscaling of Foam Flow in Highly Permeable Porous Media. *Adv Water Resour* **2020**, *146*. https://doi.org/10.1016/j.advwatres.2020.103761.

(17) Soller, R.; Koehler, S. A. Rheology of Draining Steady-State Foams. *Phy Rev E* **2009**, *80* (2). https://doi.org/10.1103/PhysRevE.80.021504.

(18) Soller, R.; Koehler, S. A. Rheology of Steady-State Draining Foams. *Phys Rev Lett* **2008**, *100* (20). https://doi.org/https://doi.org/10.1103/PhysRevLett.100.208301.

(19) Buzza D. M. A.; Lu C.-Y. D.; Cates M. E. Linear Shear Rheology of Incompressible Foams. *J phys II* **1995**, *5* (1), 37–52. https://doi.org/10.1051/jp2:1995112.

(20) Cohen-Addad, S.; Hoballah, H.; Hö, R. Viscoelastic Response of a Coarsening Foam. *Phys Rev E* **1998**, *57* (6), 6897–6901. https://doi.org/https://doi.org/10.1103/PhysRevE.57.6897.

(21) Krishan, K.; Helal, A.; Höhler, R.; Cohen-Addad, S. Fast Relaxations in Foam. *Phys Rev E* **2010**, *82* (1), 011405. https://doi.org/https://doi.org/10.1103/PhysRevE.82.011405.

(22) Gopal, A. D.; Durian, D. J. Relaxing in Foam. *Phys Rev Lett* **2003**, *91* (18). https://doi.org/https://doi.org/10.1103/PhysRevLett.91.188303.

(23) Koehler, S. A.; Hilgenfeldt, S.; Stone, H. A. A Generalized View of Foam Drainage: Experiment and Theory. *Langmuir* **2000**, *16* (15), 6327–6341. https://doi.org/10.1021/la9913147.

(24) Hilgenfeldt, S.; Koehler, S. A.; Stone, H. A. Dynamics of Coarsening Foams: Accelerated and Self-Limiting Drainage. *Phys Rev Lett* **2001**, *66* (10). https://doi.org/10.1103/PhysRevLett.86.4704.

(25) Woodward, J. G. A Vibrating-Plate Viscometer. *J Acoust Soc Am* **1953**, *25* (1), 147–151. https://doi.org/10.1121/1.1906989.

(26) Iida, T.; Kawamoto, M.; Fujimoto, S.; Morita, Z.-I. An Experimental Investigation on Characteristics of the New Oscillating-Plate Viscometer. *Tetsu-to-Hagané* **1985**, *71* (11), 1490–1496. https://doi.org/https://doi.org/10.2355/tetsutohagane1955.71.11_1490.

(27) AKPEK, A.; YOUN, C.; KAGAWA, T. A Study on Vibrational Viscometers Considering Temperature Distribution Effect. *JFPS Int J Fluid Power* **2014**, *7* (1), 1–8. https://doi.org/https://doi.org/10.5739/jfpsij.7.1.

(28) JR Calvert, K. N. Bubble Size Effects in Foams . *Int J Heat Fluid Flow* **1987**, *8* (2), 102–106. https://doi.org/https://doi.org/10.1016/0142-727X(87)90005-1.

(29) Weaire, D.; Hutzler, S. *The Physics of Foams*; Oxford University Press, 1999, P10.

(30) Osei-Bonsu, K.; Grassia, P.; Shokri, N. Relationship between Bulk Foam Stability, Surfactant Formulation and Oil Displacement Efficiency in Porous Media. *Fuel* **2017**, *203*, 403–410. https://doi.org/10.1016/j.fuel.2017.04.114.

(31) Osei-Bonsu, K.; Shokri, N.; Grassia, P. Foam Stability in the Presence and Absence of Hydrocarbons: From Bubble- to Bulk-Scale. *Colloids Surf A Physicochem Eng Asp* **2015**, *481*, 514–526. https://doi.org/https://doi.org/10.1016/j.colsurfa.2015.06.023.

(32) Yu, W.; Zhou, X.; Kanj, M. Y. Microfluidic Investigation of Foam Coarsening Dynamics in Porous Media at High-Pressure and High-Temperature Conditions. *Langmuir* **2022**, *38* (9), 2895–2905. https://doi.org/https://doi.org/10.1021/acs.langmuir.1c03301.

(33) Langevin, D. Recent Advances on Emulsion and Foam Stability. *Langmuir*. American Chemical Society March 21, 2023, pp 3821–3828. https://doi.org/10.1021/acs.langmuir.2c03423.

(34) Sheng, Y.; Lin, K.; Binks, B. P.; Ngai, T. Ultra-Stable Aqueous Foams Induced by Interfacial Co-Assembly of Highly Hydrophobic Particles and Hydrophilic Polymer. *J Colloid Interface Sci* **2020**, *579*, 628–636. https://doi.org/10.1016/j.jcis.2020.06.098.

(35) Behrens, S. H. Oil-Coated Bubbles in Particle Suspensions, Capillary Foams, and Related Opportunities in Colloidal Multiphase Systems. *Curr Opin Colloid Interface Sci* **2020**, *50*, 101384. https://doi.org/10.1016/j.cocis.2020.08.009.